\documentclass[10pt,english,aps,prd,nofootinbib,preprintnumbers]{revtex4}
\usepackage{graphicx}
\usepackage{dcolumn}
\usepackage{bm}
\usepackage{xcolor}
\usepackage[latin9]{inputenc}
\usepackage{amsmath}
\usepackage{amssymb}
\usepackage{amsfonts}
\usepackage{lscape}
\usepackage{hyperref}
\usepackage{url}
\usepackage{color}
\usepackage{tabularx}
\usepackage{mathtools}
\newcommand{\be}{\begin{equation}}
\newcommand{\ee}{\end{equation}}
\newcommand{\ba}{\begin{eqnarray}}
\newcommand{\ea}{\end{eqnarray}}
\newcommand{\nn}{\nonumber}

\renewcommand{\[}{\begin{equation}}
\renewcommand{\]}{\end{equation}}
\def\be{\begin{equation}}
\def\ee{\end{equation}}
\def\bea{\begin{eqnarray}}
\def\eea{\end{eqnarray}}
\def\eqi{\begin{equation}}
\def\eqf{\end{equation}}
\def\eqia{\begin{eqnarray}}
\def\eqfa{\end{eqnarray}}
\def\lcdm{$\Lambda$CDM }

\begin{document}
\title{The effective fluid approach for modified gravity}
\author{Rub\'{e}n Arjona}
\email{ruben.arjona@uam.es}

\affiliation{Instituto de F\'isica Te\'orica UAM-CSIC, Universidad Auton\'oma de Madrid,
Cantoblanco, 28049 Madrid, Spain}

\date{\today}

\begin{abstract}
Current and coming surveys will require sub-percent agreement in theoretical accuracy to test the different cosmological and gravity scenarios. This can be performed with Boltzmann solvers, i.e. codes that solve the linear evolution of cosmological perturbations. Given the plethora of gravity models, it is crucial to have a standardized unified way to describe all of them and take them into account in a Boltzmann code. Dark Energy (DE) and Modified Gravity (MG) models, although at a first glance quite dissimilar, are possible to unify them within the same framework. In this paper we present a scenario, based on the effective fluid approach that allows to map any modified gravity model as an effective dark energy fluid and then we show how to implement it into existing Boltzmann codes in a simple and straightforward way. This approach has also the advantage that only a handful of variables are needed to compute, i.e the equation of state $w(a)$ at the background level and the sound speed $c_s^2(a,k)$ and the anisotropic stress $\pi(a,k)$ at the linear perturbation. In particular we show that with simple modifications to the latter Cosmic Linear Anisotropy Solving System (CLASS) code, which we called EFCLASS, we provide competitive results in a much simpler and less error-prone approach in including the effects of modified gravity models. To test our modifications, we particularize the effective fluid approach to f(R) theories and a surviving class of Horndeski models, the designer Horndeski (HDES) which have a background exactly to the standard cosmological model $\Lambda$CDM.
\end{abstract}

\maketitle
\section{\label{sec:1}Introduction}
The standard cosmological model $\Lambda$CDM, which is in very good agreement with all of the current data from surveys \cite{Aghanim:2018eyx,Abbott:2017wau}, is at the moment the best phenomenological explanation for a broad-spectrum of phenomena such as the existence and structure of the cosmic microwave background, primordial nucleosynthesis, the large-scale structure (LSS) in the distribution of galaxies, flat geometry of space-time \cite{Riess:2020sih} and the current accelerating expansion of the Universe \cite{Hinshaw:2012aka,Aghanim:2018eyx,Abbott:2017wau} which has also been verified in a model independent analysis through Machine Learning algorithms \cite{Arjona:2019fwb,Arjona:2020kco}.
The free parameters of this standard cosmological model with few well tried ansatzes have been measured with high precision. Despite the great success of the $\Lambda$CDM model, which assumes General Relativity (GR) on all scales, it is however only a very good phenomenological model as its main components are either unknown or not well understood. For instance, Cold Dark Matter (CDM) has not been directly detected even though there has been made a huge effort in this research field \cite{Bertone:2016nfn}. There is also a big inconsistency between both the predicted and deduced value of the cosmological constant $\Lambda$ \cite{Weinberg:1988cp,Carroll:2000fy} and there are important discrepancies between the early and late Universe observations which might hint towards new physics \cite{Verde:2019ivm}. This has led to alternative explanations and nowadays one can find a  plethora of different tests and modifications of GR and the \lcdm model. The two principal approaches which circumvent the need of a cosmological constant are Dark Energy (DE) models \cite{Copeland:2006wr}, where the presence of scalar fields would accelerate the Universe at late times without the need of fine-tuning \cite{Ratra:1987rm}, and Modified Gravity (MG) models that instead modify the current theory of gravity, i.e GR \cite{Clifton:2011jh}.\\

A critical aspect of this attempt is to be able to accurately compute a wide range of observables from the cosmological models, where current and coming surveys such as Euclid \footnote{https://www.euclid-ec.org/}, LSST \footnote{https://www.lsst.org/}, WFIRST \footnote{https://wfirst.gsfc.nasa.gov/}, SKA \footnote{http://skatelescope.org/}, and Stage 4 CMB \footnote{https://cmb-s4.org/} experiments will require sub-percent agreement in theoretical accuracy to test these different scenarios. This can be achieved with Einstein-Boltzmann (EB) solvers, i.e. codes that solve the linear evolution of cosmological perturbations, or in other words, the linearized Einstein and Boltzmann equations on an expanding background \cite{Nadkarni-Ghosh:2016vfw}. The dynamics of cosmological perturbations, which have been extensively studied \cite{kodama1984cosmological,sugiyama1989gauge,Ma:1995ey}, are governed by the coupled Boltzmann equations for radiative and matter species and Einstein equations for the metric. There exists already publicly-available EB codes that compute the CMB polarization, temperature, and matter power spectra, e.g., the Code for Anisotropies in the Microwave Background (CAMB) \cite{Lewis:1999bs}, and the Cosmic Linear Anisotropy Solving System (CLASS) \cite{Lesgourgues:2011re} which are tested over a large range of cosmological parameters and are considered to be accurate to the sub-percent level.
Among others, these EB codes are capable to probe different gravitational theories and their cosmological consequences, test models with current data and can help in the establishment of future experiments \cite{Alonso:2016suf}. However, it is very difficult to take into account all the wide range of gravity  models at a technical level and introduce them in an EB code since each model has its own structure, equations and parameters.\\

In this paper we fill this gap and present the effective fluid approach, already studied in Refs.~\cite{Arjona:2018jhh,Arjona:2019rfn},  which allows to map any modified gravity model as an effective dark energy fluid and then we show how to implement it in the Boltzmann solver code CLASS in a simple and straightforward way finding competitive results in a much simpler and less error-prone approach. This method has also the advantage that only a handful of variables are needed to compute to describe the fluid \cite{Kunz:2012aw}, i.e the equation of state $w(a)$ at the background level and the sound speed $c_s^2(a,k)$ and the anisotropic stress $\pi(a,k)$ at the linear perturbation. Recall that for the \lcdm model ($w(a)\geq -1$, $c_s^2=0$, $\pi(a,k)=0$). This fluid approach makes easy the identification with well known single-field DE models like quintessence ($w(a)\geq -1$, $c_s^2=1$, $\pi(a,k)=0$) and K-essence ($w(a)$, $c_s^2(a)$, $\pi(a,k)=0$), where deviations from a non-zero anisotropic stress, which could be detected from weak-lensing \cite{Huterer:2010hw}, would exclude all standard DE models with a single field and would imply deviations from GR or if neglected, it could bias the cosmological parameters inferred from the data \cite{Cardona:2019qaz}.\\

It is known that both DE and MG models can fit background astrophysical observations as well as the standard cosmological model \lcdm (e.g., the so-called designer $f(R)$ models \cite{Multamaki:2005zs,delaCruzDombriz:2006fj,Pogosian:2007sw,Nesseris:2013fca}). As a consequence, these models are degenerated at the background level despite several efforts to disentangle them by using model independent approaches \cite{Nesseris:2010ep,Nesseris:2012tt}. However, the study of linear order perturbations might break this degeneracy because these models predict different growth of structures and could in principle be distinguishable from \lcdm \cite{Tsujikawa:2007gd,Pogosian:2007sw}. With the plethora of both DE and MG models it is convenient to have a unified framework which encompasses several of them. In 1974 Gregory Horndeski found the most general Lorentz-invariant extension of GR in four dimensions \cite{Horndeski:1974wa}. This theory can be obtained by using a single scalar field and restricting the equations of motion to being second order in time derivatives. The Horndeski Lagrangian comprehends theories such as Kinetic Gravity Braiding, Brans-Dicke and scalar tensor gravity, single field quintessence and K-essence theories, and $f(R)$ theories in their scalar-tensor formulation \cite{Baker:2012zs}. Although the range of models encompassed by the Horndeski Lagrangian was severely reduced (see, for instance, \cite{Creminelli:2017sry,Sakstein:2017xjx,Ezquiaga:2017ekz,Baker:2017hug,Amendola:2017orw,Crisostomi:2017pjs,
Frusciante:2018,Kase:2018aps,McManus:2016kxu,Lombriser:2015sxa,Copeland:2018yuh,Noller:2018eht,deRham:2018red}) with the recent discovery of gravitational waves by the LIGO Collaboration \cite{Abbott:2017oio}, an interesting remaining subclass of models remains including $f(R)$ theories \cite{Sotiriou:2008rp,DeFelice:2010aj,Nojiri:2017ncd,Nojiri:2010wj} and Kinetic Gravity Braiding \cite{Deffayet:2010qz}.\\

The paper is organized in the following way: In Sec. \ref{sec:2} we motivate the Effective Fluid Approach and describe its properties. In Sec.~\ref{sec:3} we briefly describe our formalism and the standard equations for perturbations in a Friedmann-Lemaitre-Robertson-Walker (FLRW) metric. In Sec.~\ref{sec:4} we discuss how f(R) models can be mapped into a DE fluid and provide analytical solutions for DE perturbations in generic f(R) models under the sub-horizon and quasi-static approximation. Then, in  Sec.~\ref{sec:5} we introduce the Horndeski Lagrangian and present a family of designer Horndeski (HDES) models. Finally, we present our modifications to the CLASS Boltzmann code and compute the CMB power spectrum for the HDES model and compare the outcome with the hi\_CLASS code in  Sec.~\ref{sec:6} and our conclusions in Sec.~\ref{sec:level4}.
\section{\label{sec:2}The Effective Fluid Approach}
Although DE and MG models are certainly stimulated by different underlying physics, it is possible to study both kind of models on the same footing. In an effective fluid approach deviations from GR can be understood as an effective fluid contribution in such a way that comparison with DE models might become relatively simple \cite{Kunz:2006ca,Pogosian:2010tj,Capozziello:2005mj,Capozziello:2006dj,Capozziello:2018ddp}. When seen as fluids, MG models can be described by an equation of state $w(a)$, a sound speed $c_s^2(a,k)$, and an anisotropic stress $\pi(a,k)$. The background is affected by the behavior of $w(a)$ while perturbations are essentially governed by $c_s^2(a,k)$ and $\pi(a,k)$. Since both DE and MG models predict different behavior for these three functions, in an effective fluid approach different models can be, to a certain degree, distinguished.\\

The main idea of the effective fluid approach is to map any MG model to an effective DE fluid by rewriting the MG theory as GR plus an effective dark energy fluid. In order to do that one must write down the equations of motion of the MG model of interest, extract the Einstein tensor $G_{\mu \nu}$ and other matter components and what remains you define it as an effective dark energy fluid through the energy momentum tensor as
\begin{equation}
\kappa T_{\mu\nu}^{(DE)}=G_{\mu \nu}- \kappa T_{\mu\nu}^{(m)},
\end{equation}
where $\kappa=8\pi G_N$ is a constant with $G_N$ being the bare Newton's constant and $T_{\mu\nu}^{(m)}$ is the energy-momentum tensor for the matter fields. The main advantages of the effective fluid approach are that it makes it easier to include in EB solver codes, since they are written as GR plus a dark energy fluid, and it provides better physical intuition as you can compute dark energy perturbation quantities both at the background and at the linear perturbation level. In what follows we will describe our theoretical framework and we will particularize the effective fluid approach to $f(R)$ theories and a surviving class of Horndeski models, the designer Horndeski (HDES) which have a background exactly to the standard cosmological model $\Lambda$CDM.

\subsection{\label{sec:3}Theoretical Framework}
To analyse the perturbations of different cosmological models we consider the Friedmann-Lemaitre-Robertson-Walker (FLRW) metric at the background level, then the perturbed FLRW metric in the conformal Newtonian gauge can be expressed as:
\be
ds^2=a(\tau)^2\left[-(1+2\Psi(\vec{x},\tau))d\tau^2+(1-2\Phi(\vec{x},\tau))d\vec{x}^2\right],
\label{eq:FRWpert}
\ee
where $\tau$ is the conformal time defined as $d\tau=dt/a(t)$ following Ref.~\cite{Ma:1995ey}.\footnote{Our conventions are: for the metric signature (-+++), the Riemann and Ricci tensors are given by $V_{b;cd}-V_{b;dc}=V_a R^a_{bcd}$ and $R_{ab}=R^s_{asb}$, and the Einstein equations are $G_{\mu\nu}=+\kappa T_{\mu\nu}$ for $\kappa=\frac{8\pi G_N}{c^4}$ where $G_N$ is the bare Newton's constant. From now on we will put the speed of light $c=1$.} We can assume an ideal fluid with an energy momentum tensor
\be
T^\mu_{\nu}=P\delta^\mu_{\nu}+(\rho+P)U^\mu U_\nu,\label{eq:enten}
\ee
where $\rho$, $P$ are the fluid density and pressure, and $U^\mu=\frac{dx^\mu}{\sqrt{-ds^2}}$ is its velocity four-vector given to first order by $U^\mu=\frac{1}{a(\tau)}\left(1-\Psi,\vec{u}\right)$, which satisfies that $U^\mu U_\mu=-1$. The elements of the energy momentum tensor to first order of perturbations are given by:
\bea
T^0_0&=&-(\bar{\rho}+\delta \rho),\nn\\
T^0_i&=&(\bar{\rho}+\bar{P})u_i,\nn\\
T^i_j&=& (\bar{P}+\delta P)\delta^i_j+\Sigma^i_j, \label{eq:effectTmn}
\eea
where $\bar{\rho},\bar{P}$ are defined on the background and are functions of time only, while the perturbations $\delta \rho, \delta P$ are functions of $(\vec{x},\tau)$ and $\Sigma^i_j\equiv T^i_j-\delta^i_j T^k_k/3$ is an anisotropic stress tensor. Then, assuming GR we find that the perturbed Einstein equations in the conformal Newtonian gauge are given by \cite{Ma:1995ey}:
\be
k^2\Phi+3\frac{\dot{a}}{a}\left(\dot{\Phi}+\frac{\dot{a}}{a}\Psi\right) = 4 \pi G_N a^2 \delta T^0_0, \label{eq:phiprimeeq}
\ee
\be
k^2\left(\dot{\Phi}+\frac{\dot{a}}{a}\Psi\right) = 4 \pi G_N a^2 (\bar{\rho}+\bar{P})\theta,\label{eq:phiprimeeq1}
\ee
\be
\ddot{\Phi}+\frac{\dot{a}}{a}(\dot{\Psi}+2\dot{\Phi})+\left(2\frac{\ddot{a}}{a}-
 \frac{\dot{a}^2}{a^2}\right)\Psi+\frac{k^2}{3}(\Phi-\Psi)
=\frac{4\pi}{3}G_N a^2\delta T^i_i,
\ee
\be
k^2(\Phi-\Psi) = 12\pi G_N a^2 (\bar{\rho}+\bar{P})\sigma \label{eq:anisoeq},
\ee
where we have defined the velocity $\theta\equiv ik^ju_j$, the anisotropic stress  $(\bar{\rho}+\bar{P})\sigma\equiv-(\hat{k}_i\hat{k}_j-\frac13 \delta_{ij})\Sigma^{ij}$. Finally, at the linear order, to describe the evolution equations for the perturbations we need two first order equations:  the evolution equations for the perturbations and the velocity of the fluid which can be derived through the energy-momentum conservation $T^{\mu\nu}_{;\nu}=0$ as
\be
\dot{\delta} = -(1+w)(\theta-3\dot{\Phi})-3\frac{\dot{a}}{a}\left(c_s^2-w\right)\delta,
\label{eq:cons1}
\ee
\be
\dot{\theta} = -\frac{\dot{a}}{a}(1-3w)\theta-\frac{\dot{w}}{1+w}\theta+\frac{c_s^2}{1+w}k^2\delta-k^2\sigma+k^2\Psi,
\label{eq:cons2}
\ee
where we define the equation of state parameter as $w\equiv\frac{\bar{P}}{\bar{\rho}}$,  the anisotropic stress parameter of the fluid as $\pi\equiv\frac32(1+w)\sigma$ and the rest-frame sound speed of the fluid as $c_s^2\equiv\frac{\delta P}{\delta \rho}$. For the velocity equation of the fluid we will use the scalar velocity perturbation $V\equiv i k_jT^j_0/\rho=(1+w)\theta$ instead of the velocity $\theta$. The former has the advantage that it can remain finite when the equation of state $w$ of the fluid crosses $-1$ (see Ref.~\cite{Sapone:2009mb}). With this new variable the evolution equations for the perturbations read
\bea
\delta' &=& 3(1+w) \Phi'-\frac{V}{a^2 H}-\frac{3}{a}\left(\frac{\delta P}{\bar{\rho}}-w\delta\right),
\label{Eq:evolution-delta}
\eea
\bea
V' &=& -(1-3w)\frac{V}{a}+\frac{k^2}{a^2 H}\frac{\delta P}{\bar{\rho}} +(1+w)\frac{k^2}{a^2 H} \Psi -\frac23 \frac{k^2}{a^2 H} \pi,
\label{Eq:evolution-V}
\eea
where the prime $'$ is a derivative with respect to the scale factor $a$ and $H(t)=\frac{da/dt}{a}$ is the Hubble parameter. From Eqs.~(\ref{Eq:evolution-delta},\ref{Eq:evolution-V}) we see that at linear order our MG model can be described in the effective fluid approach with just three variables, the equation of state $w(a)$ at the background level and the sound speed $c_s^2(a,k)$ or equivalently the pressure perturbation $\delta P(a,k)$ and the anisotropic stress $\pi(a,k)$ at the linear perturbation. Following we will show how to compute these quantities in $f(R)$ theories and in a surviving class of Horndeski's Lagrangian and explain how to introduce these quantities properly in the EB solver CLASS.

\subsection{\label{sec:4}f(R) models}
In the case of the $f(R)$ models, the modified Einstein-Hilbert action reads:
\be
S=\int d^{4}x\sqrt{-g}\left[  \frac{1}{2\kappa}f\left( R\right)
+\mathcal{L}_{m}\right],  \label{eq:action1}%
\ee
where $\mathcal{L}_{m}$ is the Lagrangian of matter and $\kappa=8\pi G_N$. Varying the action with respect to the metric, following the metric variational approach, we arrive at the following field equations \cite{Tsujikawa:2007gd}:
\be
F G_{\mu\nu}-\frac12(f(R)-R~F) g_{\mu\nu}+\left(g_{\mu\nu}\Box-\nabla_\mu\nabla_\nu\right)F =\kappa\,T_{\mu\nu}^{(m)},
\label{eq:EE}
\ee
where $F=f'(R)$. By adding and subtracting the Einstein tensor on the left hand side of Eq.~(\ref{eq:EE}) and moving everything to the right hand side we can rewrite the equations of motion as the usual Einstein equations plus an effective DE fluid, along with the usual matter fields \cite{Pogosian:2010tj}:
\bea
G_{\mu\nu}&=&\kappa\left(T_{\mu\nu}^{(m)}+T_{\mu\nu}^{(DE)}\right),
\label{eq:effEqs}
\eea
where
\bea
\kappa T_{\mu\nu}^{(DE)}=(1-F)G_{\mu\nu}+\frac12(f(R)-R~F) g_{\mu\nu} -\left(g_{\mu\nu}\Box-\nabla_\mu\nabla_\nu\right)F.
\label{eq:effTmn}
\eea
Due to the diffeomorphism invariance of the theory, it is very easy to show that the effective energy momentum tensor given by Eq.~\eqref{eq:effTmn}, indeed satisfies the usual conservation equation:
\be
\nabla^\mu T_{\mu\nu}^{(DE)}=0.
\ee
As expected, the background equations are the same as in GR \cite{Ma:1995ey}:
\bea
\mathcal{H}^2&=&\frac{\kappa}{3}a^2 \left(\bar{\rho}_{m}+\bar{\rho}_{DE}\right), \\
\dot{\mathcal{H}}&=&-\frac{\kappa}{6}a^2 \left(\left(\bar{\rho}_{m}+3\bar{P}_{m}\right)+\left(\bar{\rho}_{DE}+3\bar{P}_{DE}\right)\right).
\eea
We are assuming that the matter is pressureless ($\bar{P}_{m}=0$), then the effective DE density and pressure are given by:
\bea
\kappa \bar{P}_{DE}&=&\frac{f}2-\mathcal{H}^2/a^2-2F\mathcal{H}^2/a^2+\mathcal{H}\dot{F}/a^2 -2\dot{\mathcal{H}}/a^2-F\dot{\mathcal{H}}/a^2+\ddot{F}/a^2,\label{eq:effpr}\\
\kappa \bar{\rho}_{DE}&=&-\frac{f}2+3\mathcal{H}^2/a^2-3\mathcal{H}\dot{F}/a^2+3F\dot{\mathcal{H}}/a^2,\label{eq:effden}
\eea
where $\mathcal{H}=\frac{\dot{a}}{a}$ is the conformal Hubble parameter.\footnote{In what follows we denote the usual Hubble parameter as $H(t)=\frac{da/dt}{a}$ and the conformal one as $\mathcal{H}(\tau)=\frac{da/d\tau}{a}$. The two are related via $\mathcal{H}(\tau)=a H(t)$.} Using Eqs.~(\ref{eq:effpr}) and (\ref{eq:effden}) we see that the DE equation of state for the $f(R)$ models in the effective fluid description is given by:
\be
w_{DE}=\frac{-a^2 f+2\left((1+2F)\mathcal{H}^2-\mathcal{H}\dot{F}+(2+F)\dot{\mathcal{H}}-\ddot{F}\right)}{a^2 f-6(\mathcal{H}^2-\mathcal{H}\dot{F}+F\dot{\mathcal{H}})}\label{eq:wde},
\ee
which is in agreement with the expression found in Ref.\cite{Tsujikawa:2007gd}. With this formalism, it is clear that by working in the effective fluid approach, we can assign a density, pressure, velocity and anisotropic stress to the effective energy momentum tensor as in the general case of Eq.~(\ref{eq:effectTmn}). Then, we can find the effective quantities for the $f(R)$ model using the tensor of Eq.~(\ref{eq:effTmn}). We will compute these quantities in the sub-horizon and quasistatic approximation, the reason is due to the fact that the expressions for DE perturbations might be cumbersome. For example, the perturbation in the Ricci scalar is
\bea
\delta R&=&-\frac{12  (\mathcal{H}^2+\dot{\mathcal{H}})}{a^2}\Psi-\frac{4 k^2}{a^2}\Phi+\frac{2 k^2 }{a^2}\Psi \nn
-\frac{18 \mathcal{H} }{a^2}\dot{\Phi}-\frac{6 \mathcal{H} }{a^2}\dot{\Psi}-\frac{6 \ddot{\Phi}}{a^2}\label{eq:ricciexact}\\
&\simeq & -\frac{4 k^2}{a^2}\Phi+\frac{2 k^2}{a^2}\Psi,\label{eq:ricciapp}
\eea
where the last line follows from the sub-horizon approximation. We have explained in great detail the way we carry out the subhorizon and quasistatic approximations in our previous paper (see Sec. II.A.1 in Ref. \cite{Arjona:2018jhh}), but in short, the former refers to only considering modes deep in the Hubble radius, i.e. those for which $k^2 \gg a^2 H^2$, while the latter refers to neglecting derivatives of the potentials during matter domination as they are roughly constant but also terms of similar order as $\partial_\eta\sim 1/\eta\sim aH(a)$. As a result, the effective pressure, density and velocity perturbations are given by:
 \bea
\frac{\delta P_{DE}}{\bar{\rho}_{DE}}&\simeq&\frac{1}{3F}\frac{2\frac{k^2}{a^2}\frac{F_{,R}}{F}+3(1+5\frac{k^2}{a^2}\frac{F_{,R}}{F})\ddot{F}k^{-2}}{1+3\frac{k^2}{a^2}\frac{F_{,R}}{F}}\frac{\bar{\rho}_m}{\bar{\rho}_{DE}} \delta_m,\label{eq:effpres} \nn \\
& & \\
\delta_{DE}&\simeq&\frac{1}{F}\frac{1-F+\frac{k^2}{a^2}(2-3F)\frac{F_{,R}}{F}}{1+3\frac{k^2}{a^2}\frac{F_{,R}}{F}}\frac{\bar{\rho}_m}{\bar{\rho}_{DE}} \delta_m,\label{eq:effder} \\
V_{DE}&\equiv& (1+w_{DE})\theta_{DE} \simeq\frac{\dot{F}}{2F}\frac{1+6\frac{k^2}{a^2}\frac{F_{,R}}{F}}{1+3\frac{k^2}{a^2}\frac{F_{,R}}{F}}\frac{\bar{\rho}_m}{\bar{\rho}_{DE}} \delta_m.\label{eq:efftheta}
\eea
Finally, the DE anisotropic stress parameter $\pi_{DE}$ is given by
\bea
\pi_{DE}&=& \frac{\frac{k^2}{a^2} (\Phi-\Psi)}{\kappa~ \bar{\rho}_{DE}}\nn\\
&\simeq& \frac{1}{F}\frac{\frac{k^2}{a^2}\frac{F_{,R}}{F}}{1+3\frac{k^2}{a^2}\frac{F_{,R}}{F}}\frac{\bar{\rho}_m}{\bar{\rho}_{DE}} \delta_m
=\frac{\frac{k^2}{a^2}\frac{F_{,R}}{F}}{1-F+\frac{k^2}{a^2}(2-3F)\frac{F_{,R}}{F}}\delta_{DE}.\label{eq:effpi}
\eea
It is clear that for the \lcdm model, i.e., $f(R)=R-2\Lambda$, we have $F=1$ and $F_{,R}=0$ which implies that $w_{DE}=-1$ and $(\delta P_{DE},\delta \rho_{DE},\pi_{DE})=(0,0,0)$ as expected.

\subsection{\label{sec:5}Designer Horndeski}
In this section we present a family of designer Horndeski (HDES) models, i.e. models that have a background exactly equal to that of the \lcdm model but perturbations given by the Horndeski theory. In its complete form, Horndeski theory is defined as the most general Lorentz-invariant extension of GR in four dimensions and contains a few DE and MG models. Due to the recent discovery of gravitational waves by the LIGO Collaboration the Horndeski Lagrangian has been severely reduced. In particular, it has been shown that the constrain on the speed of GWs must satisfy \cite{Ezquiaga:2017ekz}
\bea
-3 \cdot 10^{-15} \le c_g/c-1 \le 7 \cdot 10^{-16},
\eea
which for Horndeski theories implies that
\bea
G_{4X}\approx 0, \hspace{2mm} G_5 \approx \text{constant}.
\eea
For the HDES models that we will present, we limit with the remaining parts of the Horndeski Lagrangian, namely,
\be
S[g_{\mu \nu}, \phi] = \int d^{4}x\sqrt{-g}\left[\sum^{4}_{i=2} \mathcal{L}_i\left[g_{\mu \nu},\phi\right] + \mathcal{L}_m \right],
\label{eq:action1}
\ee
where
\bea
\mathcal{L}_2&=& G_2\left(\phi,X\right) \equiv K\left(\phi,X\right),\\
\mathcal{L}_3&=&-G_3\left(\phi,X\right)\Box \phi,\\
\mathcal{L}_4&=&G_4\left(\phi\right) R.
\eea
Here $\phi$ is a scalar field, $X \equiv -\frac{1}{2}\partial_{\mu}\phi\partial^{\mu}\phi$ is a kinetic term, and $\Box \phi \equiv g^{\mu \nu}\nabla_{\mu}\nabla_{\nu}\phi$; $K$, $G_3$ and $G_4$ are free functions of $\phi$ and $X$.\footnote{From now on we define $G_i \equiv G_i\left(\phi,X\right)$, $G_{i,X} \equiv G_{iX} \equiv \frac{\partial G_i}{\partial X}$ and $G_{i,\phi} \equiv G_{i\phi} \equiv \frac{\partial G_i}{\partial \phi}$ where $i=2,3,4$.} Since we are mainly interested in the late-time dynamics of the Universe, from now on we will assume $\mathcal{L}_m$ is the Lagrangian of a CDM component. Among the theories embedded in the action \eqref{eq:action1} one finds, for example f(R) theories \cite{Chiba:2003ir}, Brans-Dicke theories,  \cite{Brans:1961sx} and Cubic Galileon \cite{Quiros:2019ktw}. An interesting DE model is the kinetic gravity braiding (KGB) which is characterized by the following Lagrangian
\begin{equation}
K=K(X), \hspace{5mm} G_3=G_3(X), \hspace{5mm} G_4=\frac{1}{2 \kappa}.\label{eq:KGB:definition}
\end{equation}
Considering Eq.~\eqref{eq:KGB:definition}, we will now describe how to find specific designer models such that the background is always that of the \lcdm model, namely, having $w_{DE}=-1$ but having different perturbations. This is particularly useful in detecting deviations from \lcdm at the perturbations level and is a natural expansion of our earlier work \cite{Nesseris:2013fca,Arjona:2018jhh}. For that purpose we need two functions, the modified Friedmann equation and the scalar field conservation equation which have been presented and discussed extensively in \cite{Arjona:2019rfn}.  We start with the modified Friedmann equation, which can be written as \bea
\label{eq:friedeq}
 -H(a)^2-\frac{K(X)}{3}+H^2_0\Omega_m(a)+2\sqrt{2}X^{3/2}H(a)G_{3X}+\frac{2}{3}X K_X=0,
\eea
where $\Omega_m(a)$ is the matter density and $H_0$ is the Hubble parameter. The scalar field conservation equation can be written as
\begin{equation}
\label{eq:scfeq}
    \frac{J_c}{a^3}-6XH(a)G_{3X}-\sqrt{2}\sqrt{X}K_X=0,
\end{equation}
where $J_c$ is a constant which quantifies our deviation from the attractor, as in the case of the KGB model \cite{Kimura:2010di}. We now have two equations given by \eqref{eq:friedeq} and \eqref{eq:scfeq}, but three unknown functions $(G_{3X}(X),K(X), H(a))$ thus the system is undetermined. Therefore, we need to specify one of the three unknown functions $(G_{3X}(X),K(X), H(a))$ and determine the other two using Eqs.~\eqref{eq:friedeq} and \eqref{eq:scfeq}. To facilitate this, we express the Hubble parameter as a function of the kinetic term $X$, ie $H=H(X)$ and then solve the previous equations to find $(G_{3X}(X),K(X))$. Doing so yields:
\bea
\label{eq:systemdes}
K(X) &=& -3 H_0^2 \Omega_{\Lambda,0}+\frac{J_c \sqrt{2X} H(X)^2}{H_0^2 \Omega_{m,0}}-\frac{J_c \sqrt{2X} \Omega_{\Lambda,0}}{\Omega_{m,0}}, \nn\\
G_{3X}(X) &=& -\frac{2 J_c H'(X)}{3 H_0^2 \Omega_{m,0}},
\eea
where $\Omega_{m,0}$ is the matter density at redshift $z=0$ (today) and $\Omega_{\Lambda,0}=1-\Omega_{m,0}$. With Eqs.~\eqref{eq:systemdes} we can make a whole family of designer models that behave like \lcdm at the background level but have different perturbations. A specific model found, which we call HDES \cite{Arjona:2019rfn}, that has a smooth limit to \lcdm and it also recovers GR when $J_c\sim0$ is the following. First, we demand that the kinetic term behaves as $X= \frac{c_0}{H(a)^n}$, where $c_0>0$ and $n>0$. Then, from Eqs.~\eqref{eq:scfeq} and \eqref{eq:friedeq} we find:
\bea
\label{eq:bestdes}
G_{3}(X)&=&-\frac{2 J_c c_0^{1/n} X^{-1/n}}{3 H_0^2 \Omega_{m,0}},\\
K(X)&=&\frac{\sqrt{2} J_c c_0^{2/n} X^{\frac{1}{2}-\frac{2}{n}}}{H_0^2 \Omega_{m,0}}-3 H_0^2 \Omega_{\Lambda,0}-\frac{\sqrt{2} J_c \sqrt{X} \Omega_{\Lambda,0}}{\Omega_{m,0}}.\nn
\eea
In the following section we will extend the effective fluid approach to this particular Horndeski model (HDES) and describe how to implement the dark energy effective formulae in the Boltzmann solver CLASS.
\section{\label{sec:6}CMB power spectrum and implementation in CLASS}
Here we present our modifications to the CLASS Boltzmann code, which we call EFCLASS. We will compare the outcome with the hi\_CLASS code, which solves the full set of dynamical equations but at the cost of significantly more complicated modifications. As mentioned in Sec. \ref{sec:2}, a fluid can be described by its equation of state, sound speed, and anisotropic stress. Following the fluid approach described for $f(R)$ theories in in Sec. \ref{sec:4}, we can define a DE effective energy-momentum tensor $T_{\mu\nu}^{DE}$ obtained via the gravitational field equations of the Horndeski Lagrangian and defined explicitly as follows:
\bea
G_{\mu\nu}&=&\kappa\left(T_{\mu\nu}^{(m)}+T_{\mu\nu}^{(DE)}\right),\nn \\
\kappa T_{\mu\nu}^{(DE)}&=&G_{\mu\nu}-2\kappa \sum^{4}_{i=2}\mathcal{G}^{i}_{\mu \nu},\label{eq:tmunude}
\eea
where the specific form of the gravitational field equations $\mathcal{G}^{i}_{\mu \nu}$ can be found at \cite{Arjona:2019rfn}. As with the $f(R)$ theory, since we are taking into consideration expressions up to linear order, $T_{\mu\nu}^{DE}$ also contains small perturbations which allow us to define quantities such as DE effective perturbations in the pressure, density, and velocity. These can be extracted from the DE effective energy-momentum tensor $T_{\mu\nu}^{DE}$ by considering the decomposition of the tensor into its components.\\

\begin{figure*}[!t]
\centering
\includegraphics[width=0.49\textwidth]{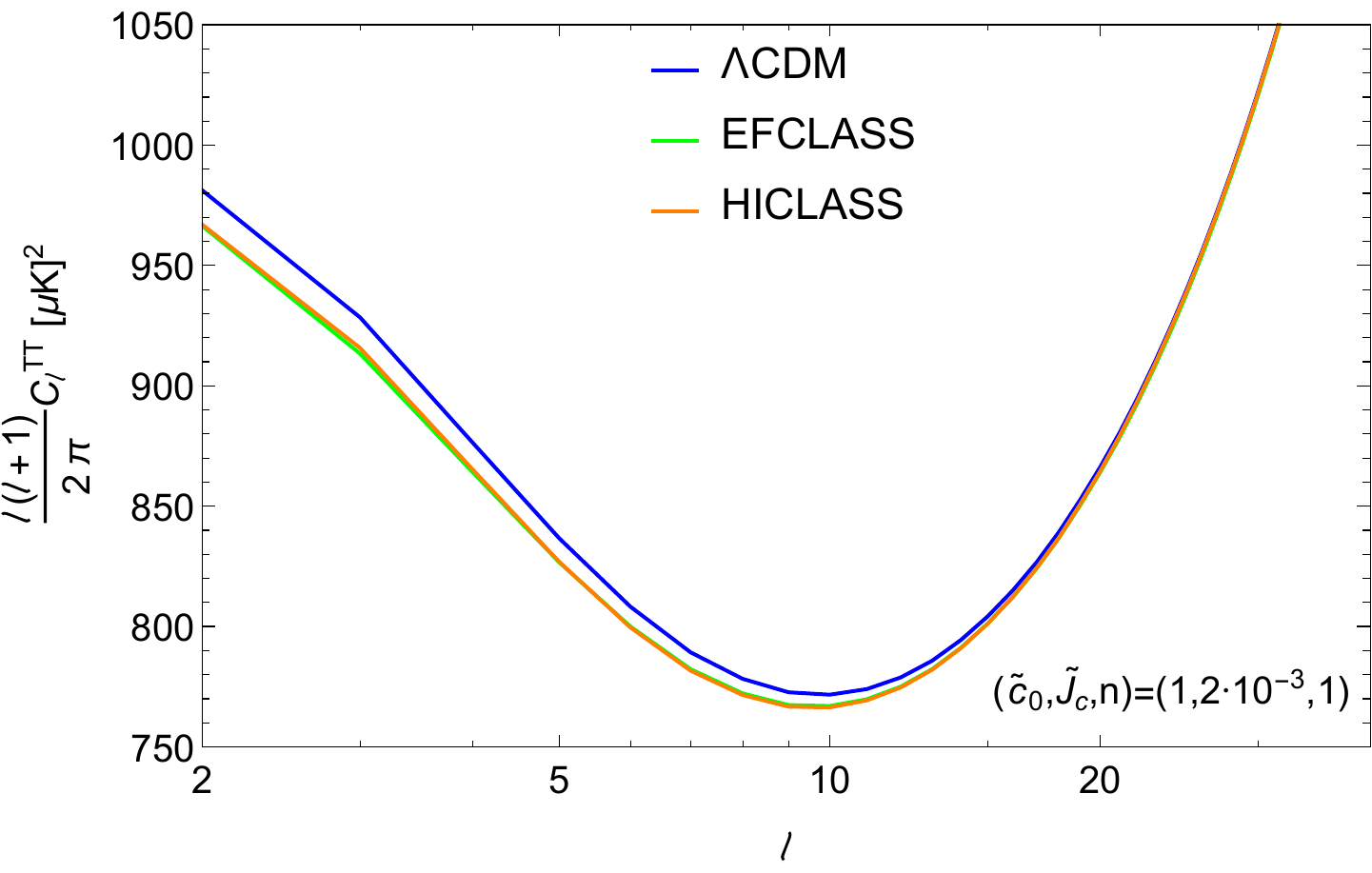}
\includegraphics[width=0.49\textwidth]{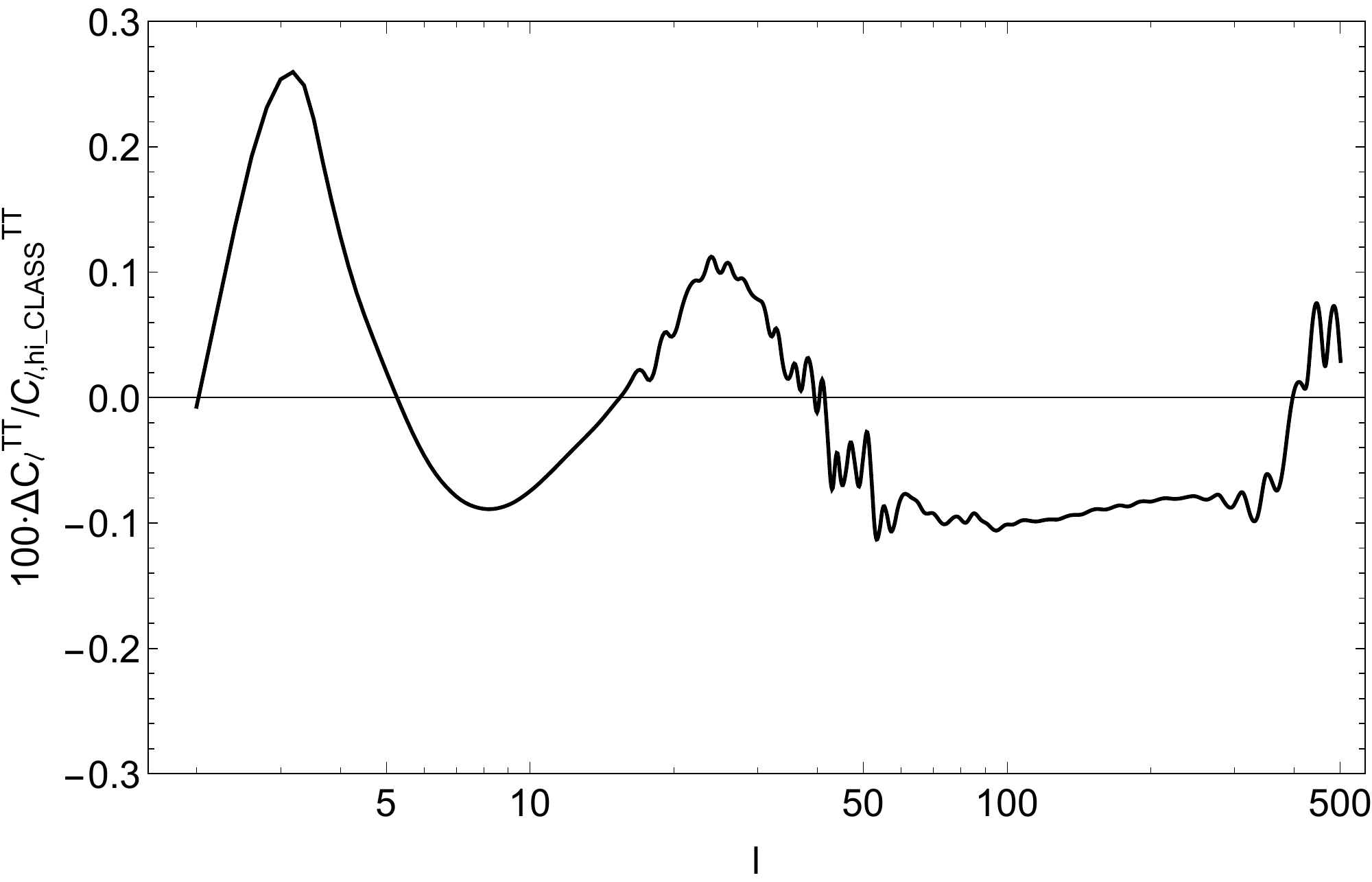}
\caption{Left: The low-l multipoles of the TT CMB spectrum for a flat universe with $\Omega_{m,0}=0.3$, $n_s=1$, $A_s=2.3 \cdot 10^{-9}$, $h=0.7$ and $(\tilde{c_0},\tilde{J_c},n)=(1,2\cdot 10^{-3},1)$. The values of values for $\tilde{J}_c$ were chosen so as to highlight the differences of these models with respect to GR.
Our EFCLASS code is denoted by the green line, hi\_CLASS by the orange line and for reference the \lcdm with a blue line. Right: The percent difference of our code with hi\_CLASS as a reference. As can be seen, our simple modification achieves roughly $\sim 0.1\%$ accuracy across all multipoles.  \label{fig:classcls}}
\end{figure*}

In order to modify the CLASS code in our effective fluid approach we only need two functions, the DE velocity and the anisotropic stress since for the HDES model the equation of state is $w_{DE}=-1$. Also for this model, because the lagrangian term $G_4$ is constant, the anisotropic stress $\pi$ is zero as can be seen from the Horndeski field equations \cite{Arjona:2019rfn}, therefore we only need the DE velocity. To have a consistent solution, we solve Eq.~\eqref{Eq:evolution-V} for the dark energy scalar velocity perturbation $V_{DE}$ and since $w_{DE}=-1$, the only variable we need is the dark energy effective pressure $\delta P_{DE}$ obtained through  Eq.~\eqref{eq:tmunude}. The expressions are rather cumbersome, but for $n=1$ we have
\be
V_{DE}\simeq \left(-\frac{14 \sqrt{2}}{3} \Omega_{m,0}^{-3/4} \tilde{J_c}~H_0~ a^{1/4}\right)\frac{\bar{\rho}_m}{\bar{\rho}_{DE}} \delta_m.\label{eq:VDEHDES}
\ee
In the left panel of Fig.~\ref{fig:classcls} we show the low-$\ell$ multipoles of the TT CMB spectrum for a flat universe with $\Omega_{m,0}=0.3$, $n_s=1$, $A_s=2.3 \cdot 10^{-9}$, $h=0.7$ and $(\tilde{c_0},\tilde{J_c},n)=(1,2\cdot 10^{-3},1)$. Our EFCLASS code is denoted by the green line, hi\_CLASS by the orange line and for reference the \lcdm with a blue line. On the right panel of Fig.~\ref{fig:classcls} we show the percent difference of our code with hi$\_$CLASS as a reference\footnote{In this case we did not use $n=2$ as we found that in this case hi\_CLASS crashes and we cannot compare with that code.}. As can be seen, our simple modification achieves roughly $\sim 0.1\%$ accuracy across all multipoles.

Finally we present our implementation of the effective fluid approach in the CLASS code \cite{Blas:2011rf}, which we call EFCLASS.
The only changes we made in the code is in the \emph{perturbations.c} file where we included the proper perturbations for the effective DE fluid given by Eqs.~(\ref{eq:phiprimeeq1}) and (\ref{eq:anisoeq}). We found that the most straight-forward and least error-prone way to make these changes is to modify the $\Lambda$CDM model equations in the aforementioned part of the code, as in the case of the perturbations, $\Lambda$CDM has none so we can just add the appropriate new terms given by Eqs.~(\ref{eq:phiprimeeq1}) and (\ref{eq:anisoeq}). We found that the best place to implement the modifications were in the \emph{perturb\_einstein} routine of CLASS, which solves the Einstein equations in the conformal Newtonian gauge given by Eqs.~(\ref{eq:phiprimeeq1}) and (\ref{eq:anisoeq}). Then, it is simple to just add in the right-hand-side of the aforementioned equations our expressions for the effective fluid DE velocity and anisotropic stress, which for the HDES model is zero.\\

Our analytic approach has several advantages: First, given that most viable $f(R)$ models and our HDES model can be written as small perturbations around $\Lambda$CDM model, it is always possible to derive extremely accurate expressions for the background, as was shown in Ref.~\cite{Basilakos:2013nfa}. Second, regarding the perturbations our improved subhorizon approximation gives much more accurate results compared to codes that are based on the default subhorizon approximation. Also, the accuracy is comparable to codes that treat the perturbations exactly by numerically solving the relevant equations. However, our approach has a much smaller overhead in terms of new lines of code and as a result is more straight-forward and less error-prone.

\section{\label{sec:level4}Conclusion}
Given the plethora of modified gravity and dark energy theories where each model has its own structure, equations and parameters it is very difficult to analyse all of them at a technical level in an Einstein Boltzmann solver code. In this paper we have presented the effective fluid approach which allows to map any modified gravity model as an effective dark energy fluid and then we show how to implement it in the Boltzmann solver code CLASS in a simple and straightforward way finding competitive results in a much simpler and less error-prone approach in including the effects of modified gravity models. This method has also the advantage that only three variables are needed to compute to describe the fluid, i.e the equation of state $w(a)$ at the background level and the sound speed $c_s^2(a,k)$ and the anisotropic stress $\pi(a,k)$ at the linear perturbation.\\

In this paper we discussed in detail the effective fluid approach and perturbation theory in the context of $f(R)$ theories. We presented several new results, in particular regarding the effective DE fluid components of the energy momentum tensor, the effective velocity of the fluid $V_{DE}$ given by Eq.~(\ref{eq:efftheta}), the effective pressure and anisotropic stress by Eqs.~(\ref{eq:effpres}) and (\ref{eq:effpi}). We then presented a family of designer Horndeski models, i.e. models that have a background exactly equal to that of the CDM model but perturbations given by the Horndeski theory. We implemented the parameterized version for the DE effective fluid of our $w_{DE}=-1$ designer Horndeski HDES model in the public code CLASS, which we call EFCLASS, by following the straightforward implementation explained with the $f(R)$ model.\\

For the sake of comparison and in order to check the validity of our effective fluid approach, we compared results from our code EFCLASS with the public code hi$\_$CLASS, which solves numerically the full perturbation equations. In Fig.~\ref{fig:classcls} we show the CMB angular power spectrum computed with both codes and as can be seen in the right panel of Fig.~\ref{fig:classcls}, the agreement is remarkable and on average on the order of $\sim0.1\%$. Since the hi$\_$CLASS code does not utilize neither the subhorizon nor the quasistatic approximation, but our EFCLASS does it, we conclude our effective fluid approach is quite accurate and powerful. Furthermore, the main advantage of our method is that while hi\_CLASS requires significant and non-trivial modifications, our EFCLASS code practically only requires the implementation of Eq.~\ref{eq:VDEHDES}, which is trivial.\\


\textbf{Numerical Analysis Files}: The numerical codes for the modifications to the CLASS code, dubbed EFCLASS, can be found at \href{https://members.ift.uam-csic.es/savvas.nesseris/efclass.html}{https://members.ift.uam-csic.es/savvas.nesseris/efclass.html}, but also at the GitHub repositories \href{https://github.com/wilmarcardonac/EFCLASS}{https://github.com/wilmarcardonac/EFCLASS} and \href{https://github.com/RubenArjona/EFCLASS}{https://github.com/RubenArjona/EFCLASS}.

\section*{Acknowledgements}
It is a pleasure to thank S. Nesseris for useful discussions. We also acknowledge support from the Research Projects FPA2015-68048-03-3P [MINECO-FEDER], PGC2018-094773-B-C32 and the Centro de Excelencia Severo Ochoa Program SEV-2016-0597.

\bibliography{effluid}

\end{document}